\documentclass[pra,twocolumn,showpacs,floatfix]{revtex4}
\usepackage{graphicx}
\begin{document}

\title{Manipulating the rotational properties of a two-component
Bose gas}
\author{J. Christensson$^1$, S. Bargi$^1$, K. K\"arkk\"ainen$^1$,
Y. Yu$^1$, G. M. Kavoulakis$^1$, M. Manninen$^2$, and S. M.
Reimann$^1$}
\affiliation{$^1$Mathematical Physics, Lund
Institute of Technology, P.O. Box 118, SE-22100 Lund, Sweden
\\
$^2$Nanoscience Center, Department of Physics, FIN-40014
University of Jyv\"askyl\"a, Finland}
\date{\today}

\begin{abstract}

A rotating, two-component Bose-Einstein condensate is shown to
exhibit vortices of multiple quantization, which are possible
due to the interatomic interactions between the two species.
Also, persistent currents are absent in this system. Finally,
the order parameter has a very simple structure for a range of
angular momenta.

\end{abstract}
 \maketitle

When a superfluid is set into rotation, it demonstrates many
fascinating phenomena, such as quantized vortex states and
persistent flow \cite{Leggett1}. The studies of rotational
properties of superfluids originated some decades ago, mostly
in connection with liquid Helium, nuclei, and neutron stars.
More recently, similar properties have also been studied
extensively in cold gases of trapped atoms.

Quantum gases of atoms provide an ideal system for studying
multi-component superfluids. At first sight, the rotational
properties of a multi-component gas may look like a trivial
generalization of the case of a single component. However, as
long as the different components interact and exchange angular
momentum, the extra degrees of freedom associated with the
motion of each species is not at all a trivial effect. On the
contrary, this coupled system may demonstrate some very
different phenomena, see, e.g., Refs.\,\cite{Ho1,Ho2,Ho3}.
Several experimental and theoretical studies have been
performed on this problem, see, e.g.,
Refs.\,\cite{Cornell1,Holland,Ketterle,Cornell2,HoMu,Ueda1,Ueda2},
as well as the review article of Ref.\,\cite{Uedareview}.

In this Letter, the rotational properties of a superfluid that
consists of two distinguishable components are examined. Three
new and surprising conclusions result from our study:

Firstly, under appropriate conditions, one may achieve vortex
states of multiple quantization. It is important to note that
these states result from the interaction between the different
species, and not from the functional form of the external
confinement. It is well known from older studies of
single-component gases, that any external potential that
increases more rapidly than quadratically gives rise to vortex
states of multiple quantization, for sufficiently weak
interactions \cite{Emil}; on the contrary, in a harmonic
potential, the vortex states are always singly-quantized. In
the present study, vortex states of multiple quantization
result purely because of the interaction between the different
components, even in a harmonic external potential. Therefore,
our study may serve as an alternative way to achieve such
states \cite{Dalibard}.

Secondly, our simulations indicate that multi-component gases
do not support persistent currents, in agreement with older
studies of homogeneous superfluids \cite{Ho2}. Essentially, the
energy barrier that separates the (metastable) state with
circulation/flow from the non-rotating state, is absent in this
case, as the numerical results, as well as the intuitive
arguments presented below, suggest.

Finally, we investigate the structure of the lowest state of
the gas, in the range of the total angular momentum $L$ between
zero and $N_{\rm min} = {\rm min}(N_A, N_B)$, where $N_A$ and
$N_B$ are the populations of the two species labelled as $A$
and $B$. In this range of $L$, only the single-particle states
with $m=0$ and $m=1$ are macroscopically occupied, as derived
in Ref.\,\cite{BCKR} within the approximation of the lowest
Landau level of weak interactions. Remarkably, our numerical
simulations within the mean-field approximation, which go well
beyond the limit of weak interactions, show that this result is
more general.

For simplicity we assume equal masses for the atoms of the two
components, $M_A = M_B = M$. Also, we model the elastic
collisions between the atoms by a contact potential, with equal
scattering lengths for collisions between the same species and
different species, $a_{AA}=a_{BB}= a_{AB}=a$ (except in
Fig.\,4). Our results are not sensitive to the above equality
and hold even if $a_{AA} \approx a_{BB} \approx a_{AB}$, as in
Rubidium, for example. For the atom populations we assume $N_A
\neq N_B$, but $N_A/N_B \alt 1$ (without loss of generality).
The trapping potential is assumed to be harmonic, $V_{\rm ext}
({\bf r}) = M (\omega^2 \rho^2 + \omega_z^2 z^2)/2$. Our
Hamiltonian is thus
\begin{eqnarray}
    {\hat H} = \sum_{i=1}^{N_A + N_B} - \frac {\hbar^2 \nabla_i^2} {2M}
  + V_{\rm ext} ({\bf r}_i)
 + \frac {U_0} 2 \sum_{i \neq j=1}^{N_A+N_B} \delta({\bf r}_i-{\bf r}_j),
\end{eqnarray}
where $U_0 = 4 \pi \hbar^2 a/M$. We consider rotation around
the $z$ axis, and also assume that $\hbar \omega_z \gg \hbar
\omega$, and $\hbar \omega_z \gg n_0 U_0$, where $n_0$ is the
typical atom density. With these assumptions, our problem
becomes effectively two-dimensional, as the atoms reside in the
lowest harmonic oscillator state along the axis of rotation.
Thus, there are only two quantum numbers that characterize the
motion of the atoms, the number of radial nodes $n$, and the
quantum number $m$ associated with the angular momentum. The
corresponding eigenstates of the harmonic oscillator in two
dimensions are labelled as $\Phi_{n,m}$.

Within the mean-field approximation, the energy of the gas in
the rest frame is
\begin{eqnarray}
    E &=& \sum_{i=A,B} \int \Psi_i^*
    \left( - \frac {\hbar^2 \nabla^2} {2M}
  + V_{\rm ext} ({\bf r}) \right) \Psi_i \, d^3r +
\nonumber \\
 &+& \frac {U_0} 2
 \int (|\Psi_A|^4 + |\Psi_B|^4 + 2 |\Psi_A|^2 |\Psi_B|^2)
\, d^3 r,
\label{en}
\end{eqnarray}
where $\Psi_A$ and $\Psi_B$ are the order parameters of the two
components. By considering variations in $\Psi_A^*$ and
$\Psi_B^*$, we get the two coupled Gross-Pitaevskii-like
equations,
\begin{eqnarray}
 \left (- \frac {\hbar^2 \nabla^2} {2M} + V_{\rm ext}
 + U_0 |\Psi_B|^2 \right) \Psi_A
 + U_0 |\Psi_A|^2 \Psi_A &=& \mu_A \Psi_A,
 \nonumber \\
 \left(- \frac {\hbar^2 \nabla^2} {2M} + V_{\rm ext}
 + U_0 |\Psi_A|^2 \right) \Psi_B
+ U_0 |\Psi_B|^2 \Psi_B  &=& \mu_B \Psi_B,
\nonumber \\
\label{GPE}
\end{eqnarray}
where $\mu_A$ and $\mu_B$ is the chemical potential of each
component. We use the method of relaxation \cite{relax} to
minimize the energy of Eq.\,(\ref{en}) in the rotating frame,
$E' = E - L \Omega$, where $\Omega$ is its angular velocity.

For the diagonalization of the many-body Hamiltonian, we
further assume weak interactions, $n_0 U_0 \ll \hbar \omega$,
and work within the subspace of the states of the lowest Landau
level, with $n = 0$. This condition is not necessary, however
it allows us to consider a relatively larger number of atoms
and higher values of the angular momentum. We consider all the
Fock states which are eigenstates of the number operators
${\hat N}_A$, ${\hat N}_B$ of each species, and of the operator
of the total angular momentum ${\hat L}$, and diagonalize the
resulting matrix.

Combination of the mean-field approximation and of numerical
diagonalization of the many-body Hamiltonian allows us to
examine both limits of weak as well as strong interactions. For
obvious reasons we use the diagonalization in the limit of weak
interactions, and the mean-field approximation (mostly) in the
limit of strong interactions. The interaction energy is
measured in units of $v_0 = U_0 \int |\Phi_{0,0}(x,y)|^4
|\phi_0(z)|^4 \, d^3r = (2/\pi)^{1/2} \hbar \omega a/a_z$,
where $\phi_{0} (z)$ is the lowest state of the oscillator
potential along the $z$ axis, and $a_z = (\hbar/M
\omega_z)^{1/2}$ is the oscillator length along this axis. For
convenience we introduce the dimensionless constant $\gamma = N
v_0/\hbar \omega = \sqrt{2/\pi} N a/a_z$, with $N = N_A + N_B$
being the total number of atoms, which measures the strength of
the interaction.

We first study the limit of weak coupling, $\gamma \ll 1$, and
use numerical diagonalization. Considering $N_A = 4$ and $N_B =
16$ atoms, we use the conditional probability distributions to
plot the density of the two components, for $L = 4, 16, 28$,
and 32, as shown in Fig.\,1. When $L = 4 = N_A$, and $L = 16 =
N_B$, the component whose population is equal to $L$ forms a
vortex state at the center of the trap, while the other
component does not rotate, residing in the core of the vortex.
This is a so-called ``coreless" vortex state
\cite{Cornell1,Ketterle,coreless}. As $L$ increases beyond $L =
N_B = 16$, a second vortex enters component $B$, and for $L = 2
N_B = 32$, this merges with the other vortex to form a
doubly-quantized vortex state. For this value of $L = 32$, the
smaller component $A$ does not carry any angular momentum
(apart from corrections of order $1/N$). The fact that this is
indeed a doubly-quantized vortex state, is confirmed by the
occupancy of the single-particle states. By increasing $N_A,
N_B$, and $L = 2N_B$ proportionally, we observe that the
occupancy of the single-particle state with $m=2$ of component
$B$ approaches unity, while the occupancy of all the other
states are at most of order $1/N_B$. The same happens for the
single-particle state with $m=0$ of the non-rotating component
$A$.

A similar situation emerges for the case of stronger coupling,
$\gamma = 50$, where we have minimized the mean-field energy of
Eq.\,(\ref{en}) in the rotating frame (in the absence of
rotation the two clouds do not phase separate). For example, we
get convergent solutions, shown in Fig.\,2, for $N_B/N_A =
2.777$, and (i): $L_A = N_A, L_B = 0$, for $\Omega/\omega =
0.35$ (top left), (ii) $L_A = 0, L_B = N_B$, for $\Omega/\omega
= 0.45$ (top middle), (iii) $L_A = 0.755 N_A, L_B = 1.171 N_B$,
for $\Omega/\omega = 0.555$ (top right), (iv) $L_A = 0, L_B = 2
N_B$, for $\Omega/\omega = 0.60$ (bottom left), (v) $L_A =
0.876 N_A, L_B = 2.057 N_B$, for $\Omega/\omega = 0.69$ (bottom
middle), and (vii) $L_A = 0, L_B = 3 N_B$, for $\Omega/\omega =
0.73$ (bottom right). Here, $L_A$ and $L_B$ is the angular
momentum of each component, with $L = L_A + L_B$. Again, when
$L = 2 N_B$, and $L = 3 N_B$ the phase plots show clearly a
doubly-quantized and a triply-quantized vortex state in
component $B$, and a non-rotating cloud in component $A$.

\begin{figure}[t]
\includegraphics[width=6.5cm,height=3.1cm,angle=-0]{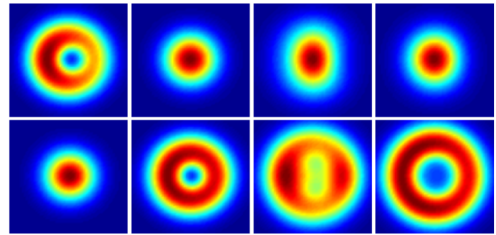}
\caption[]{The conditional probability distribution of the two
components, with $N_A = 4$ (higher row), and $N_B = 16$ (lower
row). Each graph extends between $-2.4 a_0$ and $2.4 a_0$ in
both directions. The reference point is located at $(x,y) =
(1.25 a_0, 0)$ in the lower graphs ($B$ component). The angular
momentum $L$ increases from left to right, $L = 4 (= N_A), 16
(= N_B)$, 28, and $32 (= 2N_B)$.}
\label{FIG1}
\end{figure}

The picture that appears from these calculations is intriguing:
as $\Omega$ increases, a multiply-quantized vortex state of
multiplicity $\kappa$ splits into $\kappa$ singly-quantized
ones, and on the same time, one more singly-quantized vortex
state enters the cloud from infinity. Eventually all these
vortices merge into a multiply-quantized one of multiplicity
equal to $\kappa+1$. Figure 2 shows the above results for
various values of $\Omega$.

The mechanical stability of states which involve the gradual
entry of the vortices from the periphery of the cloud is novel.
This behavior is absent in one-component systems, in both
harmonic, as well as anharmonic trapping potentials. In one
component gases, only vortex phases of given rotational
symmetry are mechanically stable \cite{RokhsarB,KB}. In the
present problem, the mechanical stability of states with no
rotational symmetry (shown in Fig.\,2) is a consequence of the
non-negative curvature of the dispersion relation (i.e., of the
total energy) ${\cal E}(L)$. This observation also connects
with the (absence of) metastable, persistent currents (i.e.,
the second main result of our study), which we present below.

\begin{figure}[t]
\includegraphics[width=8.5cm,height=5cm,angle=0]{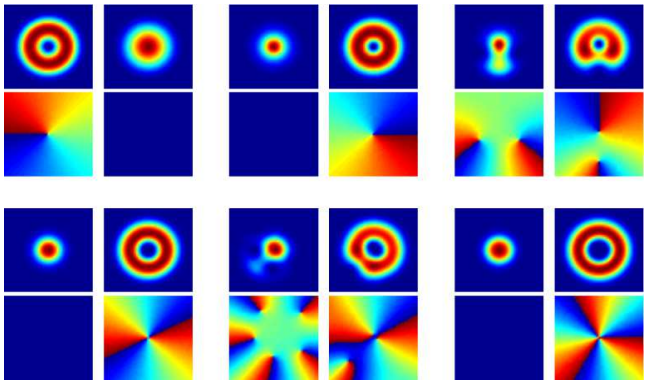}
\caption[]{The density (higher graphs of each panel) and the
phase (lower graphs of each panel) of the order parameters
$\Psi_A$ (left graphs of each panel) and $\Psi_B$  (right
graphs of each panel), with $N_B/N_A = 2.777$ and a coupling
$\gamma = 50$. Each graph extends between $-4.41 a_0$ and $4.41
a_0$ in both directions. The values of the angular momentum per
atom and of $\Omega$ in each panel are given in the text.}
\label{FIG2}
\end{figure}

In Ref.\,\cite{BCKR} we have given a simple argument for the
presence of vortex states of multiple quantization within the
mean-field approximation. At least when the ratio between $N_A$
and $N_B$ is of order unity (but $N_A \neq N_B$), there are
self-consistent solutions of Eqs.\,(\ref{GPE}) of vortex states
of multiple quantization. Within these solutions, the smaller
component (say component $A$), does not rotate, providing an
``effective" external potential $V_{{\rm eff},B}({\bf r}) =
V_{\rm ext}({\bf r}) + U_0 n_A({\bf r})$ for the other one
(component $B$), which is anharmonic close to the center of the
trap. This effectively anharmonic potential is responsible for
the multiple quantization of the vortex states. Therefore, we
conclude that for a relatively small population imbalance, the
``coreless vortices" are vortices of multiple quantization.

\begin{figure}[t]
\includegraphics[width=8cm,height=4.cm]{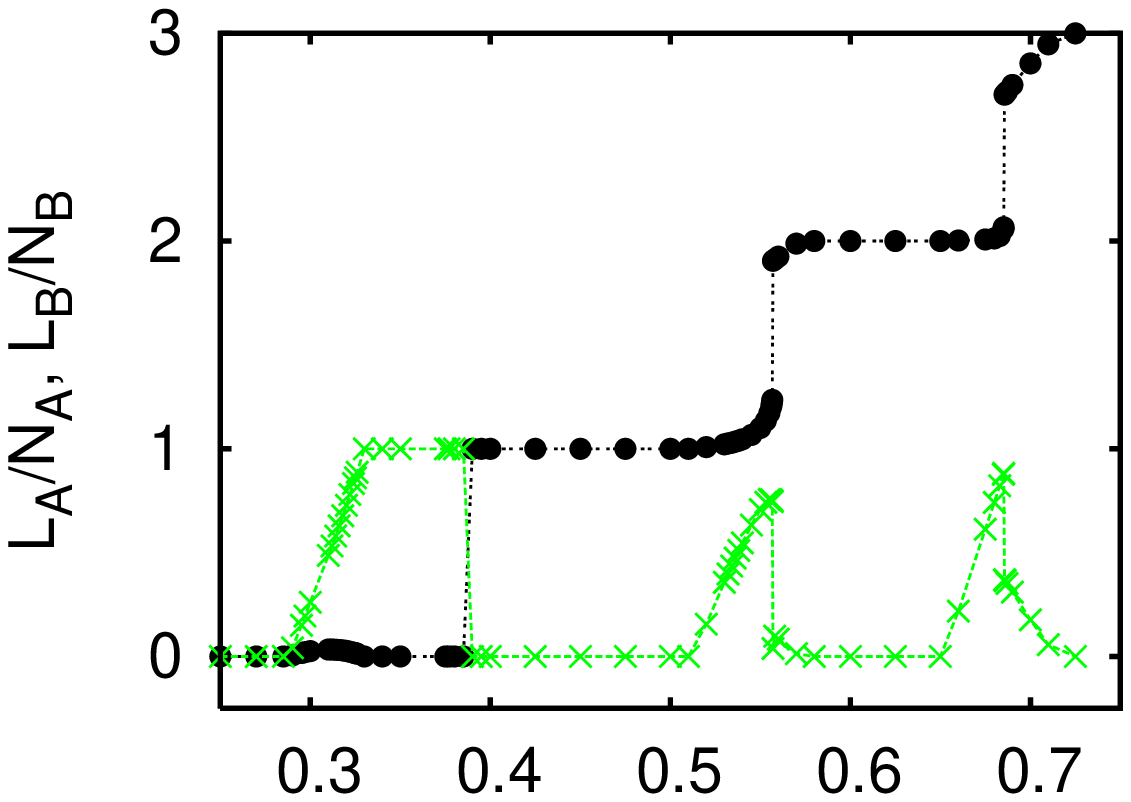}
\includegraphics[width=8cm,height=4.cm]{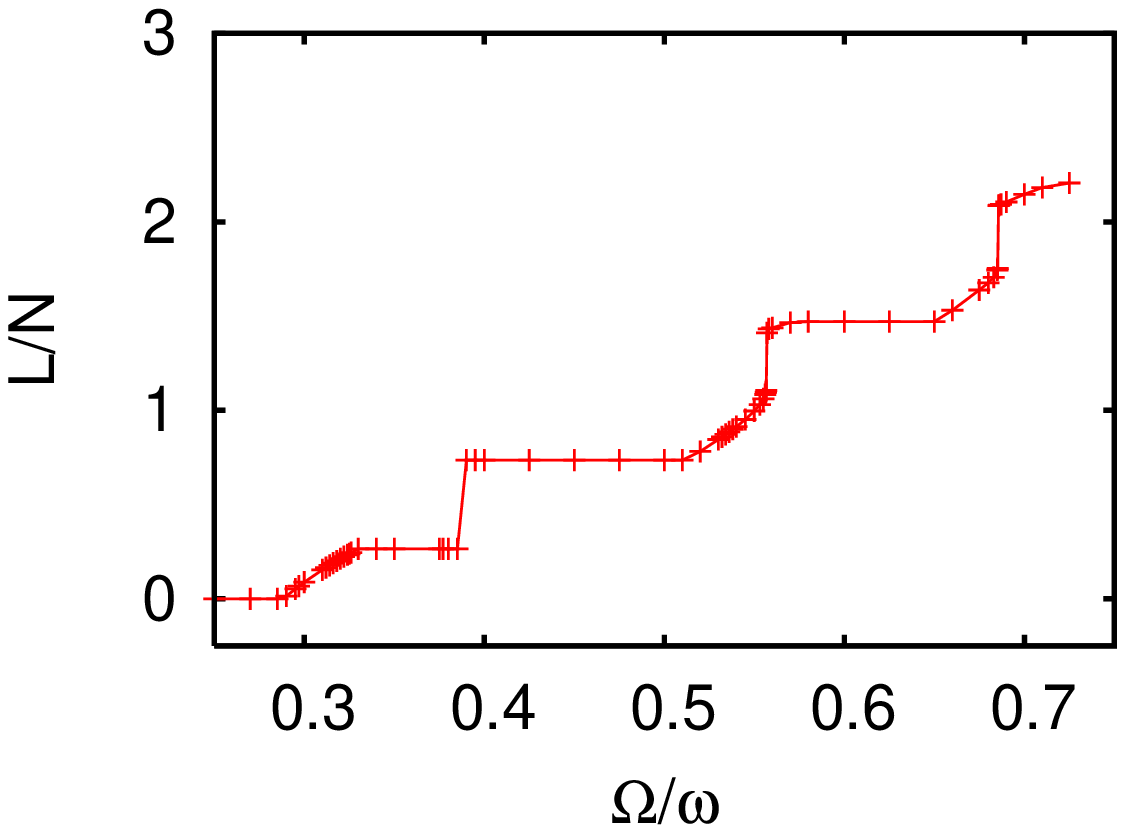}
\caption[]{Higher graph: The angular momentum $L_A/N_A$
(crosses) and $L_B/N_B$ (dots), as function of
$\Omega$. Lower graph: $L/N$ as function of $\Omega$. All
curves result from the minimization of the energy in the
rotating frame, within the mean-field approximation, for
$\gamma = 50$.}
\label{FIG3}
\end{figure}

The second aspect of our study is the absence of metastable
currents (in the laboratory frame, for $\Omega = 0$). A
convenient and physically-transparent way to think about
persistent currents is that they correspond to metastable
minima in the dispersion relation ${\cal E} = {\cal E}(L)$
\cite{Leggett}. A non-negative curvature of ${\cal E}(L)$ for
all values of $L$ implies the absence of metastability. For all
the couplings we have examined, both within the numerical
diagonalization, as well as the mean-field approximation, we
have found a non-negative second derivative of the dispersion
relation. Figure 3 shows $L_A/N_A$, $L_B/N_B$, and $L/N$ versus
$\Omega$, for $\gamma = 50$. These curves are calculated by
minimizing the energy ${\cal E}(L)$ in the rotating frame for a
fixed $\Omega$, and plotting the angular momentum per particle
of the corresponding state for the given rotational frequency.

Again, our argument for the effective anharmonic potential is
consistent with this positive curvature. Let us consider for
simplicity $a_{AA} = a_{BB} = 0$, and $a_{AB} \neq 0$. Then,
the problem of solving Eqs.\,(3) becomes essentially a
(coupled) eigenvalue problem. If $E_{0,m}$ are the (lowest)
eigenvalues of the effective (anharmonic) potential felt by the
rotating component for a given angular momentum $m \hbar$, then
$\partial^2 E_{0,m}/\partial m^2$ is always positive. For
example, if one considers a weakly-anharmonic effective
potential, $V_{\rm eff}(\rho) = M \omega^2 \rho^2 [1+\lambda
(\rho/a_0)^{2k}]/2$, where $k = 1, 2, \dots$ is a positive
integer, $a_0 = (\hbar/M \omega)^{1/2}$ is the oscillator
length, and $0 < \lambda \ll 1$ is a small dimensionless
constant, according to perturbation theory, $E_{0,m} = \hbar
\omega |m| + \lambda (|m|+1) \dots (|m|+k)/2$, which clearly
has a positive curvature.

One may gain some physical insight into the absence of
persistent currents by understanding the difference between a
gas with one and two components. In the case of a single
component, for sufficiently strong (and repulsive)
interactions, an energy barrier that separates the state with
circulation from the vortex-free state may develop. In the
simplest model where the atoms rotate in a toroidal trap, in
order for them to get rid of the circulation, they have to form
a node in their density, which costs interaction energy, and
this creates the energy barrier \cite{Leggett,Rokhsar}. On the
other hand, in the presence of a second component, this node
may be filled with atoms of the other species, and therefore
the system may get rid of the circulation with no energy
expense. This physical picture is also supported by the density
plots in Figs.\,2 and 4. For example, in the case of coreless
vortices, the core of the vortex is filled by the other
(non-rotating) component \cite{Uedareview}. More generally, the
density minima of the one component coincide, roughly speaking,
with the density maxima of the other component, resulting in a
total density $n_{\rm tot} = |\Psi_A|^2+|\Psi_B|^2$ which does
not have any local minima or nodes.

\begin{figure}[t]
\includegraphics[width=6cm,height=2.5cm,angle=0]{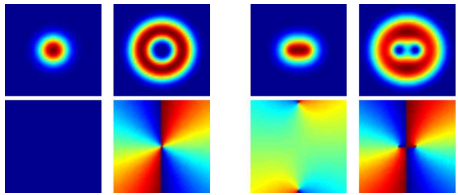}
\caption[]{The density (higher graphs) and the phase (lower
graphs) of the order parameters $\Psi_A$ (left graphs of each
panel) and $\Psi_B$ (right graphs of each panel), with $N_B/N_A
= 2.777$. Here $\Omega/\omega = 0.6$ and $\gamma = 50$. In the
left panel $L_A = 0$, and $L_B/N_B = 2$. In the right panel,
the scattering length $a_{BB}$ is twice as large as in the left
panel, $a_{BB}= 2 a$. In this case, $L_A /N_A = 0.05$, and
$L_B/N_B = 1.936$. All graphs extend between $-4.41 a_0$ and
$4.41 a_0$.}
\label{FIG4}
\end{figure}

Our third result is based on the mean-field approximation. For
$0 \le L \le N_{\rm min}$, where $N_{\rm min} = {\rm min}(N_A,
N_B)$, the only components of the order parameters $\Psi_A$ and
$\Psi_B$ are the single-particle states with $m=0$ and $m=1$,
i.e.,
\begin{eqnarray}
  \Psi_A &=& \sum_{n} c_{n,0} \Phi_{n,0} + c_{n,1} \Phi_{n,1},
\nonumber \\
  \Psi_B &=& \sum_{n} d_{n,0} \Phi_{n,0} + d_{n,1} \Phi_{n,1},
\end{eqnarray}
where $c_{n,0}, c_{n,1}, d_{n,0}$ and $d_{n,1}$ are functions
of $L$ and of the coupling. The numerical simulations that we
perform within a range of couplings $\gamma \le 50$ that extend
well beyond the lowest-Landau level approximation, reveal this
very simple structure for the lowest state of both components.
Also, the corresponding dispersion relation is numerically very
close to a parabola, as in the case of weak interactions
\cite{BCKR}. Again, one may attribute these facts to the
effective potential that arises from the interaction between
the two species \cite{BCKR}.

In the studies that have examined a single-component gas in an
external anharmonic potential, it has been shown that as the
strength of the interaction increases, there is a phase
transition from the phase of multiple quantization to the phase
of single quantization \cite{KB}. In the present case the
situation is more complex, since the effective anharmonic
potential is generated by the interaction between the two
species as a result of a self-consistent solution. Still, a
similar phase transition takes place here, as, for example, one
keeps the scattering lengths $a_{AA}$ and $a_{AB}$ fixed, and
increases $a_{BB}$ that corresponds to the rotating component.
Figure 4 shows the density and the phase of both species,
for $a_{AA} = a_{BB} = a_{AB} = a$ (left panel), and $a_{BB}
= 2 a_{AA} = 2 a_{AB} = 2 a$ (right panel). Component $B$
undergoes a phase transition from a doubly-quantized vortex
state to two singly-quantized vortices.

To conclude, mixtures of bosons demonstrate numerous novel
superfluid properties and provide a model system for studying
them. Here we have given a flavor of the richness of this
problem. Many of the results presented in our study are worth
investigating further, as, for example, one changes the ratio
of the populations, the coupling constant between the same and
different species, or the masses.

We acknowledge financial support from the European Community
project ULTRA-1D (NMP4-CT-2003-505457), the Swedish Research
Council, the Swedish Foundation for Strategic Research, and the
NordForsk Nordic Network on ``Low-dimensional physics".

\end{document}